\documentclass{pasj01}
\usepackage{lineno}

\Received{2021/12/08}
\Accepted{2022/06/07}
 
\begin{document} 
\title{Physics of nova outbursts: A theoretical model of classical nova 
outbursts with self-consistent wind mass loss
}

\author{Mariko \textsc{Kato}\altaffilmark{1}}
\altaffiltext{1}{Department of Astronomy, Keio University, Hiyoshi, Yokohama
  223-8521, Japan}
\email{mariko.kato@hc.st.keio.ac.jp}

\author{Hideyuki \textsc{Saio},\altaffilmark{2}}
\altaffiltext{2}{Astronomical Institute, Graduate School of Science,
    Tohoku University, Sendai, 980-8578, Japan}

\author{Izumi \textsc{Hachisu}\altaffilmark{3}}
\altaffiltext{3}{Department of Earth Science and Astronomy, College of Arts and
Sciences, The University of Tokyo, 3-8-1 Komaba, Meguro-ku, Tokyo 153-8902, Japan}

\maketitle
\begin{abstract}
We present a model for one cycle of a classical nova outburst based on 
a self-consistent wind mass loss accelerated by 
the gradient of radiation pressure, 
i.e., the so-called optically thick winds.  
Evolution models are calculated by a Henyey code 
for a 1.0 $M_\odot$ white dwarf (WD) 
with a mass accretion rate of $5 \times 10^{-9}~M_\odot$~yr$^{-1}$.
The outermost part of hydrogen-rich envelope is connected 
to a steadily moving envelope when optically thick winds occur. 
We confirm that no internal shock waves occur
at the thermonuclear runaway.  
The wind mass loss rate reaches a peak of 
$1.4 \times 10^{-4}~M_\odot$~yr$^{-1}$ 
at the epoch of the maximum photospheric expansion,
where the photospheric temperature decreases to $\log T_{\rm ph}$ (K)=3.90. 
Almost all of the accreted mass is lost in the wind. 
The nuclear energy generated in hydrogen burning is lost in a form of 
photon emission (64 \%), gravitational energy (lifting-up 
the wind matter against the gravity, 35 \%), and kinetic energy 
of the wind (0.23 \%). 
A classical nova should be 
very bright in a far-UV (100 - 300 \AA) band, 
during a day just after the onset of thermonuclear 
runaway ($\sim$25 d before the optical maximum). 
In the decay phase of the nova outburst, the envelope structure is very close 
to that of a steady state solution.
\end{abstract}



\section{Introduction}\label{introduction}

A nova is a thermonuclear runaway event on a mass-accreting white dwarf (WD)  
\citep{nar80,ibe82,pri86,sio79,spa78}.
During a nova outburst the hydrogen-rich envelope expands 
to a giant size and wind mass loss is accelerated. 
Such a nova outburst has been calculated 
by several groups with Henyey-type evolution codes
\citep{pri95,kov98,den13,ida13,wol13,kat14shn,kat15sh,tan14}.
These codes, however, encounter numerical difficulties 
when the nova envelope expands and wind mass loss occurs. 
To continue the numerical calculation beyond the extended stage,
various mass loss schemes are assumed. 
However, some of these mass loss rates are not always consistent
with the physical mechanism of wind acceleration
(see \citet{kat17palermo} for a review and comparison
of various numerical methods).

The most important mass-loss mechanism during a nova outburst is 
the wind accelerated by the radiation pressure-gradient deep inside
the photosphere \citep{fri66,fin71,zyt72,rug79}. 
Massive winds occur when the envelope expands and hence 
the temperature in the envelope decreases to the range 
where the radiative 
opacity increases outward \citep{kat83,kat85,kat97,kat94h,kat09a}. 
Once the optically thick winds occur, other acceleration mechanisms
such as a line-driven wind are inefficient mainly because of 
lack of photon momentum (see sections 5 and 6.2 in \citet{kat94h}). 
In order to follow the decay phase of nova outbursts  
\citet{kat83} and \citet{kat94h} developed the optically thick wind theory 
assuming that the envelope has settled to a steady state and 
thermal equilibrium. 
A sequence of steady state solutions can successfully follow the evolution
of a nova and reproduce its multiwavelength light curves 
in the decline phase. 
  
The optically thick wind theory, 
however, cannot be applied to the rising phase of a nova outburst 
because the expanding envelope has not yet reached the steady state. 
In order to follow a cycle of nova evolution, it is necessary 
to incorporate optically thick winds into evolution calculations.
Recently, \citet{kat17} succeeded in calculating multi cycles of 
{\it recurrent}  nova evolutions with a Henyey-type code including 
the optically thick winds. 
Their models are for two cases of recurrent novae, i.e., 
$1.2~M_\odot$ WD and $1.38~M_\odot$ WD with the 
mass accretion rates of $2.0 \times 10^{-7}~M_\odot$~yr$^{-1}$ 
and $1.6 \times 10^{-7}~M_\odot$~yr$^{-1}$, respectively. 
The hydrogen shell flashes in these recurrent novae are weaker
compared with those in classical novae with smaller accretion rates.

In this paper we obtain evolution models of a {\it classical} nova cycle.
The thermonuclear runaway in a classical nova is much stronger,
because the hydrogen-rich envelope accumulated before 
the ignition is more massive so that the envelope expands to a giant size. 
For this reason, modeling the evolution of a classical nova 
cycle is more demanding than the case of recurrent novae. 
We have calculated $1.0~M_\odot$ WD evolution models for one cycle 
of a classical nova outburst, assuming a quiescent accretion 
rate of $5 \times 10^{-9}~M_\odot$~yr$^{-1}$. 
Mass loss rates in expanded phases 
are set to be consistent with the optically thick wind model.  
We present the temporal change of internal structure of the envelope, 
occurrence of the wind mass loss, energy balance of mass-losing envelope, 
and ejecta mass etc. 
We also compare our results with 
multiwavelength light curves obtained by \citet{hil14}
based on the hydrogen shell flash models in \citet{pri95} and \citet{yar05}.

Because in this work we assume spherical symmetry, non-spherical phenomena
are not included.  Our model also
doesn't include physical processes that occur
outside the photosphere, e.g.,
free-free emission from ejecta \citep{hac15k},
dust formation (e.g., OS And: \citet{kik88}, V2362 Cyg \citet{ara10},
FH Ser: \citet{geh88}),
shock wave when the ejecta collide with circumbinary matter
and associated phenomena like heating, gamma ray
emission, etc. \citep{ayd20,cho21,dra10,hes22}.

This paper is organized as follows. 
Section \ref{sec_method} describes our calculation method. 
Our numerical results are presented separately, 
for one cycle of a nova outburst evolution in section \ref{sec_results}, 
temporal changes of internal structures in section \ref{sec_structure}, 
energy budget in section \ref{sec_energy}, and UV/X-ray flashes 
in section \ref{sec_UV}. 
Discussion and conclusions follow in sections \ref{sec_discussion}
and \ref{section_conclusion}, respectively.

\section{Numerical method}\label{sec_method}

We have calculated multicycle nova evolution
of a $1.0\,M_\odot$ white dwarf (WD) accreting matter 
at a rate of $5\times10^{-9}M_\odot$\,yr$^{-1}$.
We adopt this mass accretion rate as a central value of the distribution
of mass accretion rates from \citet{sel19}'s Figure 3.  Their median value
is about $3\times 10^{-9} M_\odot$\,yr$^{-1}$, which is close to our 
adopted value.
The chemical compositions of the accreting matter and initial hydrogen-rich
envelope of the WD are assumed to be $X=0.7, ~Y=0.28$, and $Z=0.02$.
We assume that the CO core is composed of 48\% of $^{12}$C, 50\% of $^{16}$O,
and 2\% of $^{20}$Ne by weight. This assumption does not
affect our results because the CO core material
hardly mixes with the upper part and the newly accreted matter
is burned into heavy elements and accumulates on the surface of CO core.
We used the same computer code with the OPAL opacity tables 
\citep{igl96} as those used 
in our models of recurrent novae published in 
\citet{kat17,kat171500,kat18hvf}, in which more
details of the method of calculations and input physics are discussed
(e.g., the number of meshes, numerical convergence, equations of state,
and nuclear burning networks).

For the initial WD model of our evolution calculations, 
we adopted a steady-state equilibrium model \citep{kat14shn}, in which 
the heating by the mass accretion and nuclear energy generation is balanced
with the cooling by radiative diffusion and neutrino emission.   
The model has parameters of $\log L/L_\odot = 2.567$, 
$\log T_{\rm c}$ (K)$=7.687$, and $\log T_{\rm ph}$ (K)$= 5.422$.
Starting from the steady-state model,
the flash properties reach a limit cycle before 5th flash cycle,
which indicates that the interior structure of our initial model is 
already a good approximation to the structure realized after many flashes.
Note that our initial WD temperature $T_{\rm c}$ is higher 
than that ($< 10^7$K) of the equilibrium core temperature of \citet{tb04}.

The equilibrium $1.0\,M_\odot$ WD model at $\dot M = 5 \times 10^{-9}~ 
M_\odot$yr$^{-1}$ is thermally unstable.  Once we start the evolution 
with mass accretion, the hydrogen shell burning weakens 
and the luminosity begins to decrease. Meanwhile, a hydrogen-rich 
envelope accumulates. When the mass of hydrogen-rich envelope becomes 
large enough (i.e. larger than the ignition mass $M_{\rm ig}$), 
the nuclear burning is re-ignited at the bottom of the envelope, 
and the envelope expands exceeding the equilibrium radius; i.e., 
a hydrogen shell flash occurs.
When the radius becomes very large (the radiation pressure 
becomes much larger than the gas pressure), the evolution calculation 
by our evolution code based on the Henyey method fails. 
Then, we assume the mass loss at a rate according to the prescription 
given in \citet{kat17}. After a certain amount of mass is lost, 
the envelope begins to contract. The mass loss is stopped 
when the radius becomes small enough \citep{kat17}.

For each stage with massloss, we calculated steady-state wind solutions with 
a different set of parameters (photospheric temperature, luminosity, 
and mass). The mass-loss rate of the wind solution is automatically determined 
from the boundary condition of the wind \citep{kat94h}.
If the wind mass-loss rate of steady-state solution does not match to the 
temporally assumed value for the Henyey code, 
we recalculate the nova evolution with different temporal mass-loss rates.

\begin{figure}
  \includegraphics[width=8cm]{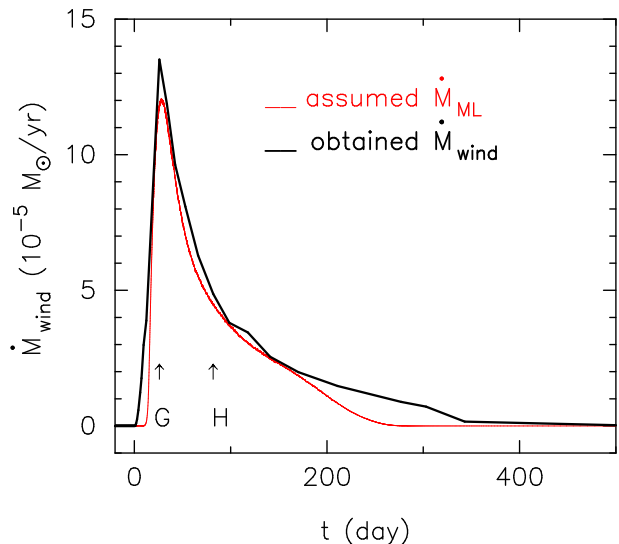}
\caption{Comparison of the temporarily assumed mass loss rates (thin red line) 
with the obtained wind mass loss rates (thick black line) after fitting. 
}\label{dmdt}
\end{figure}

Figure \ref{dmdt} shows two mass loss rates, one is the mass loss rate
initially assumed in our Henyey code and the other is the wind mass loss rate
finally obtained after fitting with the optically thick wind solutions. 
These two mass loss rates are in good agreement with each other, 
suggesting that the physical properties of our wind solutions
are reasonable (see also \citet{kat17}). 
We need many iterations until we obtain mass loss rate 
assumed for the  evolution 
model sufficiently close to the resultant wind 
mass-loss rate as described in \citet{kat17}.

\begin{longtable}{*{9}{c}}
 \caption{Characteristic properties of a shell flash model}\label{table.models}
  \hline              
      $M_{\rm WD}$ & $\dot M_{\rm acc}$ &$\log T_{\rm WD}$ & $P_{\rm rec}$&$M_{\rm acc}$ &$M_{\rm ig}$ &$\log T^{\rm max}$ & $\log L_{\rm nuc}^{\rm max}$   \\ 
     ($M_\odot$)&($M_\odot$~yr$^{-1}$)& ({\rm K}) &(yr)&($M_\odot$) &($M_\odot$) &  ({\rm K}) & ($L_\odot$) \\
\endfirsthead
  \hline
  Name & Value1 & Value2 & Value3 \\
\endhead
  \hline
\endfoot
  \hline
\endlastfoot
  \hline
     1.0  &$5 \times 10^{-9}$ &7.687 &5370  &$2.68\times 10^{-5}$ &$3.04 \times 10^{-5}$ &8.18  &8.36\\
\end{longtable}


The model parameters are summarized in Table \ref{table.models}. 
We adopt a $1.0 ~M_\odot$ WD and the mass accretion rate of 
$\dot M_{\rm acc}=5 \times 10^{-9}M_\odot$~yr$^{-1}$. We adopt 
this mass accretion rate as a central value of the distribution of
mass accretion rates obtained by \citet{sel19} as already mentioned
in section 1
(see also, e.g., Figure 6 in \citet{hac20mmrd}). 
The recurrence period is 5370 years. 
We assume that the WD interior structure 
has already reached thermal equilibrium,  
so that its central temperature $T_{\rm WD}$ is not 
a free parameter \citep{kat21}. 
The other quantities shown in Table \ref{table.models} are 
the mass accreted in the quiescent phase
$M_{\rm acc}$, the ignition mass $M_{\rm ig}$, the maximum temperature 
in the hydrogen nuclear burning layer through a flash, 
and the maximum value of nuclear luminosity. 
The difference between the accreted mass, 
$M_{\rm acc}=2.68 \times 10^{-5}~M_\odot$, and the
ignition mass, $M_{\rm ig}=3.05 \times 10^{-5}~M_\odot$, 
corresponds to the leftover from the previous outburst;
When the latest outburst ends, the hydrogen-rich envelope of 
$0.37 \times 10^{-5}~M_\odot$ remains on the WD. 
The mass accretion restarts and hydrogen-rich matter accumulates on it. 

We do not include mixing processes between the core and envelope 
prior to the shell flash, or multi-dimensional mixing 
\citep{cas10,gla12,jos20}, 
to avoid free parameters of its efficiency and to simplify our calculations.  
After the onset of thermonuclear runaway, an extensive convection zone
appears and the processed helium is carried upward to reduce 
the hydrogen content $X$ in the envelope, from 0.7 to 0.65, 
whereas heavy element content $Z=0.02$ is unchanged. 
Effects of these simplification are discussed
in subsection \ref{sec_composition}.  




\section{One cycle of shell flashes}\label{sec_results}

\subsection{H-R diagram}\label{sec_hr}

\begin{figure}
  \includegraphics[width=8cm]{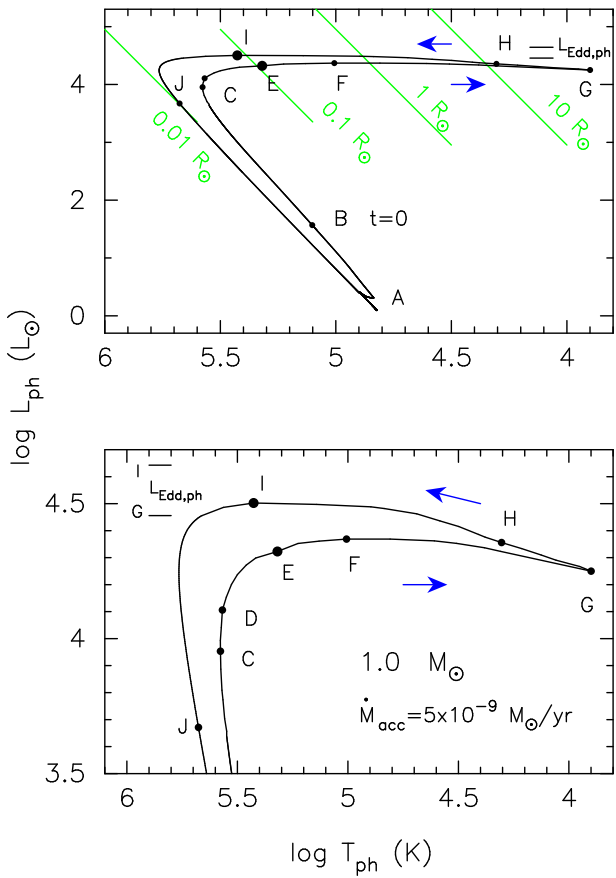}
\caption{The H-R diagram of a shell flash on a $1.0~M_\odot$ WD with the 
mass accretion rate of $\dot M_{\rm acc}=5 \times 10^{-9}M_\odot$~yr$^{-1}$.
Lower panel is an enlargement of the top part of the upper panel. 
A: Quiescent phase before the shell flash. 
B: The epoch when $L_{\rm nuc}$ reaches the maximum value ($t=0$).
C: The photospheric temperature takes the maximum value in the rising phase. 
D: X-ray flash (see section \ref{sec_UV}). 
E: The wind mass loss begins. 
F: The epoch at $\log T_{\rm ph}$ (K)=5.0. 
G: The maximum expansion when the wind mass loss rate and photospheric 
radius reach maximum.
H: The epoch at $\log T_{\rm ph}$ (K)=4.3.  
I: The wind mass loss stops.
J: The supersoft X-ray luminosity decreases to one tenth of the maximum value. 
}\label{m10.hr}
\end{figure}

\begin{longtable}{*{10}{c}}
 \caption{Time since the beginning of the flash}\label{table.time}
  \hline              
      Stage & B & C & D & E & F & G & H & I & J \\
      & $L_{\rm nuc}^{\rm max}$ &$T_{\rm ph}^{\rm max}$ &X flash  & wind starts  && max exp && wind ends
       & $0.1~L_{\rm X}^{\rm max}$\\
\endfirsthead
  \hline
\endhead
  \hline
\endlastfoot
  \hline
     Time (day)  & 0.0 & 0.063 & 0.114 & 1.05 & 2.83&26.0&81.9 & 530&6.46 yr \\
$\log T_{\rm ph}$~({\rm K}) &5.102 &5.577& 5.568&5.319 &5.006& 3.900&4.305&5.427& 5.676\\
$\dot M_{\rm wind}(10^{-5}M_\odot$~yr$^{-1}$)&  -  & - &-  & 0.0 &0.25&13.5&4.9&0.0&- \\
\end{longtable}


Figure \ref{m10.hr} shows the evolutionary track 
in the H-R diagram for one cycle of a classical nova outburst.
In the quiescent phase the accreting WD stays 
around the bottom of the line (denoted by label A). 
After the thermonuclear runaway sets in, the luminosity 
 increases with keeping the photospheric radius almost constant. 
We define the onset of the shell flash 
at the time when $L_{\rm nuc}$ reaches maximum, the point of which is
denoted by label B ($t=0$). The photospheric temperature also increases 
and reaches maximum $T_{\rm ph}^{\rm max}$ at epoch C. 
Then, the envelope begins to expand, and 
the optically thick winds start at point E.
At stage G the photospheric radius attains its maximum and the wind 
mass loss rate also reaches maximum.
As the envelope mass decreases mainly due to wind mass loss 
and lesser extent by hydrogen burning, 
the photospheric radius decreases and the temperature
$T_{\rm ph}$ increases with time.
The winds stop at point I. After stage I 
the envelope mass continuously decreases owing
to hydrogen burning. Epoch J is the stage when the supersoft 
X-ray luminosity decreases to one tenth of its maximum value
(section \ref{sec_UV}). 
The time of each epoch is summarized in table \ref{table.time}.

Figure \ref{m10.hr} also shows two local Eddington luminosities
at the photosphere for two stages G and I,  
\begin{equation}
L_{\rm Edd,ph} \equiv {4\pi cG{M_{\rm WD}} \over\kappa_{\rm ph}},
\label{equation_Edd,ph}
\end{equation}
where $\kappa_{\rm ph}$ is the opacity at the photosphere. 
At stage G, 
$L_{\rm Edd,ph} =1.1\times 10^{38}{\rm erg~sec}^{-1}
= 2.9 \times 10^4~L_\odot$ 
($\kappa_{\rm ph}=0.458$), while at stage I, 
$L_{\rm Edd,ph} =1.7\times 10^{38}{\rm erg~sec}^{-1} 
= 4.3 \times 10^4~L_\odot$  
($\kappa_{\rm ph}=0.301$). 
These values are indicated by the short horizontal bars
in figure \ref{m10.hr}. 
The photospheric luminosity $L_{\rm ph}$
does not exceed the photospheric
Eddington luminosity $L_{\rm Edd,ph}$. 

Observationally nova brightness often exceeds 
the Eddington limit ($M_V \sim -6$, see e.g. \citet{hac04k}). 
\citet{hac06kb} explained that the observed brightness is 
owing to free-free emission originated 
from ejecta outside the photosphere, and not always attributed 
to the photospheric blackbody emission. 
The brightness of free-free emission in novae is mainly determined
from the wind mass loss rate. When the mass loss rate is larger, 
the free-free emission can exceed the Eddington limit. 
Hachisu and his group reproduced multiwavelength light curves
of a number of novae including super Eddington phase   
(\citet{hac06kb,hac08kc,hac15k,hac16a, hac18k, hac18kb, hac19ka, hac19kb,
hac20mmrd, hac21ka}). 
The sub-Eddington value of the photospheric luminosity 
does not conflict with the observational super-Eddington luminosities.

\subsection{Energy generation in thermonuclear runaway}

\begin{figure}
  \includegraphics[width=8cm]{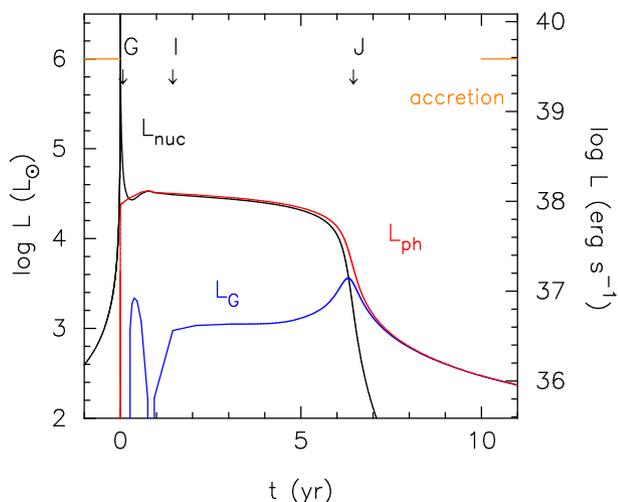}
\caption{The evolution of the photospheric luminosity, $L_{\rm ph}$,
total nuclear burning energy release rate, $L_{\rm nuc}$,
and total gravitational energy release rate, $L_{\rm G}$. 
We stopped the mass accretion at $t=7 \times 10^{-4}$ year and restarted 
at $ t=9.99$ year, the period of which is indicated with the 
horizontal orange line. 
Epoch B  in Fig.\,\ref{m10.hr} corresponds to the time 
of $t=0$ and three epochs 
G (maximum photospheric expansion), I (end of winds),
and J (supersoft X-ray luminosity decreases to one tenth of its maximum)
are indicated by the downward arrows.  
}\label{evol}
\end{figure}

Figure \ref{evol} shows the evolution of the 
emergent bolometric luminosity 
$L_{\rm ph}$, total nuclear energy release rate 
$L_{\rm nuc}=\int \epsilon_{\rm nuc} \delta m$, 
and integrated gravitational energy release rate
$L_{\rm G}=\int \epsilon_{\rm g} \delta m$. 
Here, $\epsilon_{\rm nuc}$ and $\epsilon_{\rm g}$ are energy generation 
rates per unit mass owing to nuclear burning and gravitational
energy release, respectively. 
We stopped the 
mass accretion when $L_{\rm ph}$ increases to 
$\log L_{\rm ph}/L_\odot=3.5$ ($\log T_{\rm ph}$ (K)= 5.53) 
and resumed it when the luminosity decreases to $\log L_{\rm ph}/L_\odot=2.48$ 
($\log T_{\rm ph}$ (K)$= 5.40$). The accretion phase is 
indicated with the horizontal orange lines.

In the very beginning of the outburst 
the nuclear energy generation rate $L_{\rm nuc}$ is much 
larger than the photospheric luminosity $L_{\rm ph}$. 
The rest energy is absorbed ($L_{\rm G} < 0$) and used to expand the
envelope against the gravity.
As the envelope expands, $L_{\rm nuc}$ 
decreases rapidly and becomes comparable with
$L_{\rm ph}$ (figure \ref{evol}). 
After that the absorbed energy is gradually released
as the expanded layers slowly contracts. 

\begin{figure}
  \includegraphics[width=8cm]{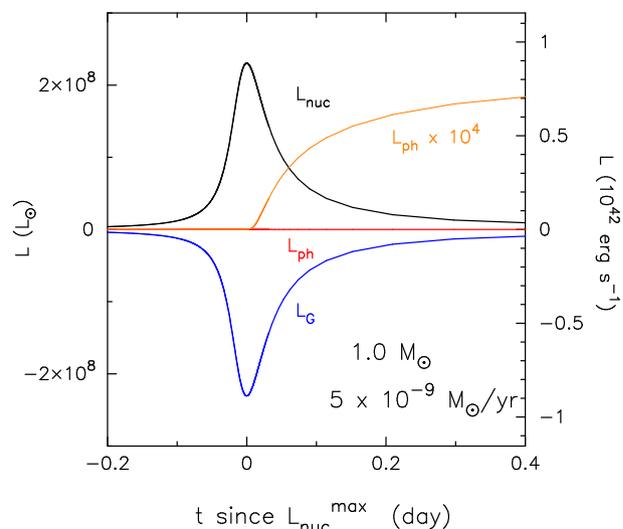}
\caption{Close-up view of a very early phase of the shell flash in
figure \ref{evol}. 
The orange line denotes $10^4$ times the photospheric luminosity 
(red line),
}\label{Levol}
\end{figure}

Figure \ref{Levol} shows a close-up view of the energy conservation during 
the epoch of thermonuclear runaway. 
The nuclear energy release rate increases in a timescale of 
$\sim 0.03$ day 
($\sim 2600$ sec) to reach the maximum value of 
$L_{\rm nuc}^{\rm max}= 2.3\times 10^{8}~L_\odot$, 
while most of the released nuclear energy is absorbed  
(i.e., $L_{\rm nuc}\sim-L_{\rm G}\gg L_{\rm ph}$) to expand the envelope.

In stage B ($t=0$) the temperature reaches $\log T$ (K) =8.18,  
where the density is $\log \rho$ (g/cm$^3$)=2.41 and 
radius is $\log R/R_\odot = -2.099$.
The local thermal timescale is roughly estimated to be
($u/\epsilon_{\rm nuc}=$) 131 sec, 
where $u$ is the specific thermal energy, 
$u=4.16 \times 10^{16}$ erg~g$^{-1}$, 
and $\epsilon_{\rm nuc}$ is the nuclear burning rate par unit mass 
$\epsilon_{\rm nuc}=3.18 \times 10^{14}$ erg g$^{-1}$sec$^{-1}$.
The dynamical timescale is obtained to be (the
pressure scale hight/local sound speed =
$5.17 \times 10^7$ cm/$1.32 \times 10^8$ cm~sec$^{-1}$=) 0.39 sec.
This timescale is much shorter than the local thermal timescale, 
which indicates that the thermonuclear runaway occurs 
nearly in hydrostatic equilibrium. 
Thus, no shock occurs.  
If the metallicity $Z$ were enhanced in the quiescent phase 
by a mixing with core material, the nuclear burning rate 
$\epsilon_{\rm nuc}$ increases but only by a factor of 4 \citep{guo22}. 
Then, no shock wave is expected in the nuclear burning zone. 
It should be noted that no internal shocks have been
reported in numerical calculations after 1990's opacity revisions.

\subsection{Degeneracy of the envelope}

\begin{figure}
  \includegraphics[width=8cm]{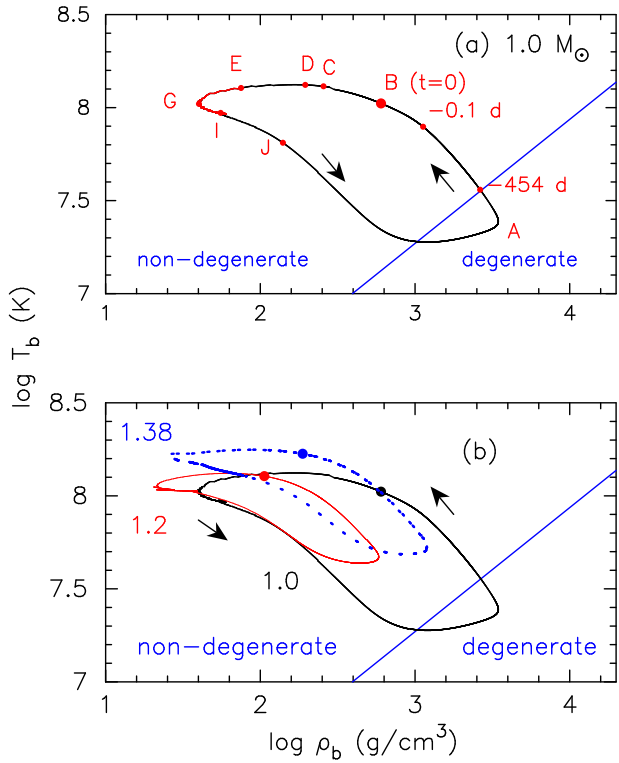}
\caption{(a) The locus of the temperature and density 
at the bottom of the hydrogen-rich envelope. 
Several characteristic stages and their corresponding time
(in units of days) are indicated. 
The blue solid straight line indicates the boundary of degenerate
and non-degenerate matter (corresponding to the degeneracy $\psi=0$). 
(b) Same as the upper panel but for a comparison with the recurrent nova
models of a $1.2~M_\odot$ WD with $\dot M_{\rm acc}=2\times 
10^{-7}M_\odot$~yr$^{-1}$ (thin red line) and a $1.38~M_\odot$ WD with
$\dot M_{\rm acc}=1.6\times 10^{-7}M_\odot$yr$^{-1}$ 
(dotted blue line) \citep{kat17}. The filled circles indicate
the stage of maximum nuclear burning rate $L_{\rm nuc}^{\rm max}$, 
corresponding to stage B in the upper panel. 
}\label{Trho}
\end{figure}

Figure \ref{Trho}(a) shows the locus of the temperature 
and density at 
the bottom of hydrogen-rich envelope 
in the $\log\rho_{\rm b}$ - $\log~T_{\rm b}$ plane 
for one cycle of a
hydrogen shell flash. 
We indicate several characteristic epochs and their times 
elapsed from  stage B ($t=0$).  
In the late accreting stage approaching stage A, 
the bottom of envelope is degenerated, 
and the density increases with little increase of temperature. 
As nuclear burning ignites at stage A, 
$T_{\rm b}$ increases by the heating, and $\rho_{\rm b}$ 
decreases by expansion, 
so that $(\rho_{\rm b}, T_{\rm b})$ 
moves upward and the degeneracy is lifted-up before 
it reaches stage B ($t=0$), where the nuclear energy generation rate
reaches its maximum. 
As nuclear burning produces a large amount of thermal energy, 
the temperature quickly increases that makes the envelope expand  
and the density decrease. 
When  $(\rho_{\rm b}, T_{\rm b})$ 
passes point B, the total nuclear burning rate 
begins to decrease because the density decreases. 
The thermonuclear flash is stabilized and 
the envelope settles to a thermal equilibrium state. 

Figure \ref{Trho}(b) shows a comparison with two recurrent nova models 
of a $1.2~M_\odot$ WD having
$\dot M_{\rm acc}=2\times 10^{-7}M_\odot$~yr$^{-1}$
and a $1.38~M_\odot$ WD having
$\dot M_{\rm acc}=1.6\times 10^{-7}M_\odot$~yr$^{-1}$
\citep{kat17}. In such high mass accretion rates, the 
envelope is hot with large gravitational energy release 
in the quiescent phase. Thus, the bottom of the envelope is 
non-degenerate. 
This is a remarkable contrast to our classical nova model
($1.0~M_\odot$ WD, $\dot M_{\rm acc}=5 \times 10^{-9}M_\odot$~yr$^{-1}$).

\section{Evolution of internal structure -- expansion and mass loss}
\label{sec_structure}

\subsection{Occurrence of optically thick winds}\label{sec_occurrence}

The optically thick wind is accelerated when the photospheric luminosity
$L_{\rm ph}$ becomes very close to the local Eddington luminosity
at the photosphere 
$L_{\rm Edd,ph}$ (Eq.\ref{equation_Edd,ph}) 
 \citep{kat85,kat09a}. 
We adopt the surface boundary condition BC1 
of \citet{kat94h}.
The occurrence condition of winds is satisfied 
 at stage E (see Figure \ref{m10.hr}), 
where the mass-loss by optically thick winds begins.
The critical point of solar-wind type solution \citep{bon52,kat94h} appears at the 
photosphere and moves inward in the mass coordinate.

\begin{figure}
  \begin{center}
  \includegraphics[width=8cm]{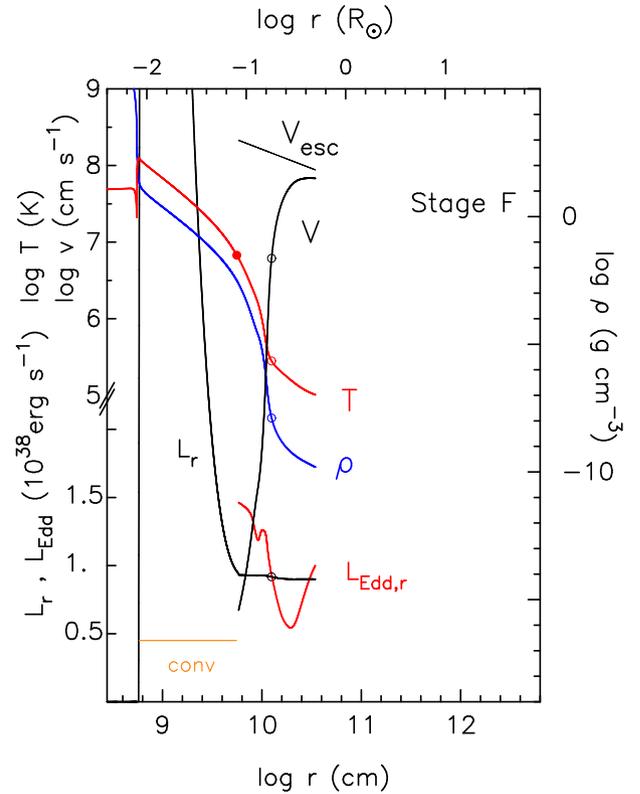}
  \end{center}
\caption{The internal structure at stage F (see Figure \ref{m10.hr}). 
The temperature, density, 
photon flux, velocity, local Eddington luminosity (lower red line), 
and escape velocity (uppermost black) are shown. The last three are plotted 
only in the outside of the fitting point.   
Convective region is denoted by the horizontal orange line. 
The open circles denote the critical point of the optically thick 
wind solution \citep{kat94h}, whereas the filled circle is the fitting point 
of our wind solution to the internal structure obtained by our Henyey code.
}\label{m10.struc.61131}
\end{figure}

Figure \ref{m10.struc.61131} shows the envelope structure at epoch F 
($\log T_{\rm ph}$ (K)=5.01), shortly after epoch E (where the wind began). 
The critical point appears in the wind acceleration 
region where the velocity (density) quickly increases (decreases). 
At this stage the wind velocity barely reaches the escape velocity
$v_{\rm esc}=\sqrt {2 G M_{\rm WD}/r}$.

This figure also shows local luminosity 
$L_{r}$, the sum of radiative and convective luminosities. 
It increases in the nuclear burning region and decreases outward 
being absorbed into the envelope. 
In the outermost layers, $L_{r}$ is almost constant.

We plot the local Eddington luminosity in the radiative 
region, which is defined as 
\begin{equation}
L_{{\rm Edd},r} \equiv {4\pi cG{M_{\rm WD}} \over\kappa}. 
\label{equation_Edd}
\end{equation}
The local Eddington luminosity is inversely proportional to the opacity 
that is a function of the temperature and density. 
A small dip at $\log r$ (cm) $\sim 9.9$ corresponds to  
a small peak in the opacity contributed by ionized O and Ne
($\log T$ (K) $\sim 6.2$--6.3), and a large dip 
at  $\log r$ (cm) $\sim 10.3$ is caused by 
a large Fe peak at $\log T$ (K) $\sim 5.2$ 
(see figure 6 in \citet{kat16xflash}). 
The wind is accelerated where the opacity increases {\it outward} 
and $L_{r}$ exceeds the local Eddington luminosity $L_{{\rm Edd},r}$.

\begin{figure}
  \begin{center}
  \includegraphics[width=8cm]{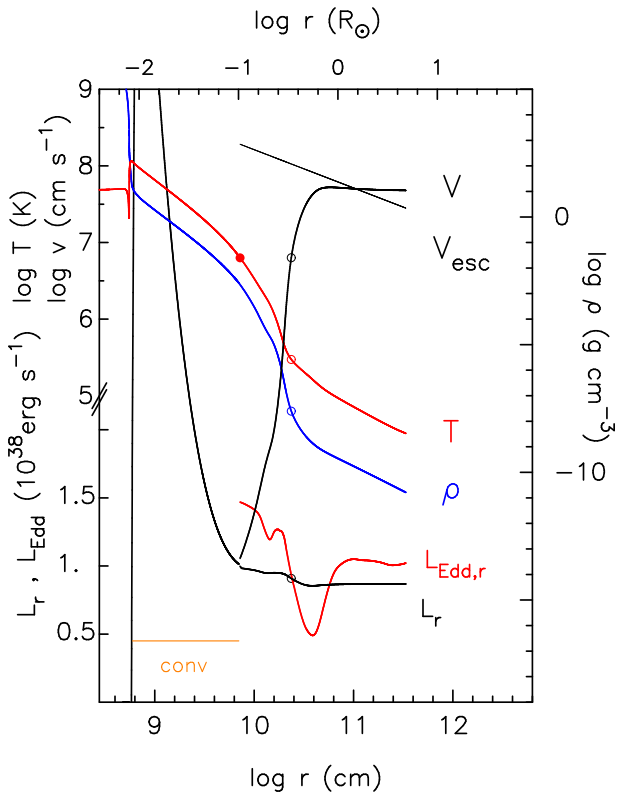}
  \end{center}
\caption{Same as in figure \ref{m10.struc.61131} 
but at a slightly later stage ($\log T_{\rm ph}$ (K)=4.51).
}\label{m10.struc.61161}
\end{figure}

Figure \ref{m10.struc.61161} shows the structure at a more expanded stage  
at $\log T_{\rm ph}$ (K)=4.51. 
The velocity reaches a constant value deep inside the photosphere and 
exceeds the escape velocity slightly below the photosphere. 
A strong acceleration occurs around the critical point 
at $\log T$ (K)=5.47 
where the opacity increases outward and the local Eddington luminosity 
quickly decreases. 
Note that the photospheric luminosity $L_{\rm ph}$ does not exceed
the photospheric Eddington luminosity $L_{\rm Edd,ph}$
whereas the local Eddington luminosity
has a dip  caused by the Fe-opacity peak 
around $\log r$ (cm) $\sim 10.6$. 

\subsection{Maximum expansion of photosphere (stage G)}

\begin{figure}
  \begin{center}
  \includegraphics[width=8cm]{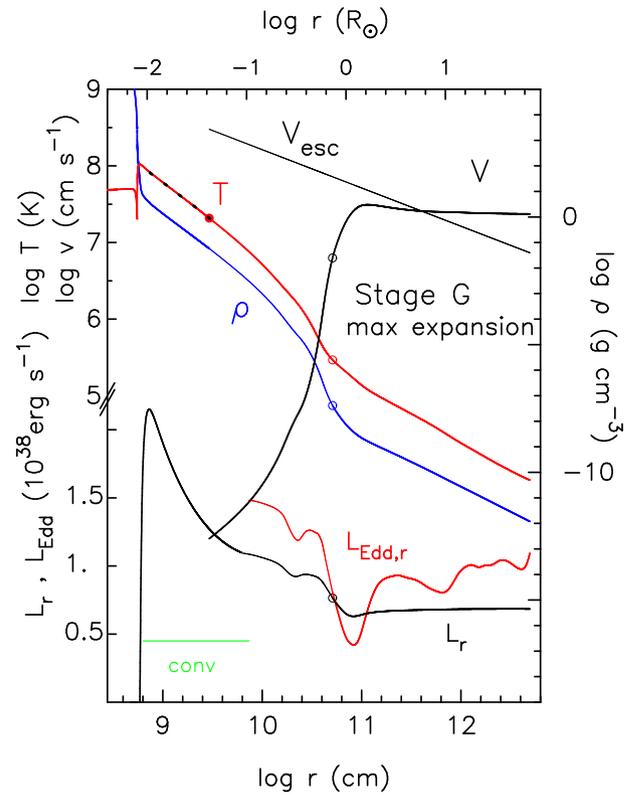}
  \end{center}
\caption{Internal structure of the envelope at the maximum expansion of the
photosphere, stage G. The temperature profile of steady state solution is 
additionally indicated beyond leftside of the fitting point 
(dotted black line), that is overlapped with the Henyey solution.    
}\label{m10.struc.max}
\end{figure}

Figure \ref{m10.struc.max} shows the distribution of the temperature, 
density, velocity and energy flux at stage G. 
The temperature and density decreases as $\rho \propto T^3$.  
The radiation pressure is much larger than the gas pressure
as $1-\beta \sim 0.8$ and $\gtsim 0.9$ 
below and above the critical point, respectively,
where $\beta \equiv P_{\rm gas}/P_{\rm total}$ 
is the ratio of the gas pressure to the total pressure.
The velocity quickly increases outward 
where the local Eddington luminosity decreases corresponding to 
the Fe peak in the opacity (at $\log T$ (K) $\sim 5.2$) and the velocity
reaches the terminal velocity deep inside the photosphere. 
The local luminosity $L_r$ decreases outward 
in the envelope below the critical point.  This is because the photon 
energy is consumed partly to lift the envelope matter up against the gravity
(gravitational energy), to heat the envelope (thermal energy),
and to increase the kinetic energy of the winds
(see figure \ref{ent} below).

The fitting point is indicated by a filled circle
on the temperature line (filled red point in figure \ref{m10.struc.max}). 
A dotted line in this figure shows the inward extension of 
the steady-state wind solution, which agrees well with 
the temperature variation in the evolution model. 
This good agreement shows that the envelope 
has settled to steady state in the almost whole envelope 
close to the nuclear burning region. 


\subsection{Decay phase}

\begin{figure}
  \begin{center}
  \includegraphics[width=8cm]{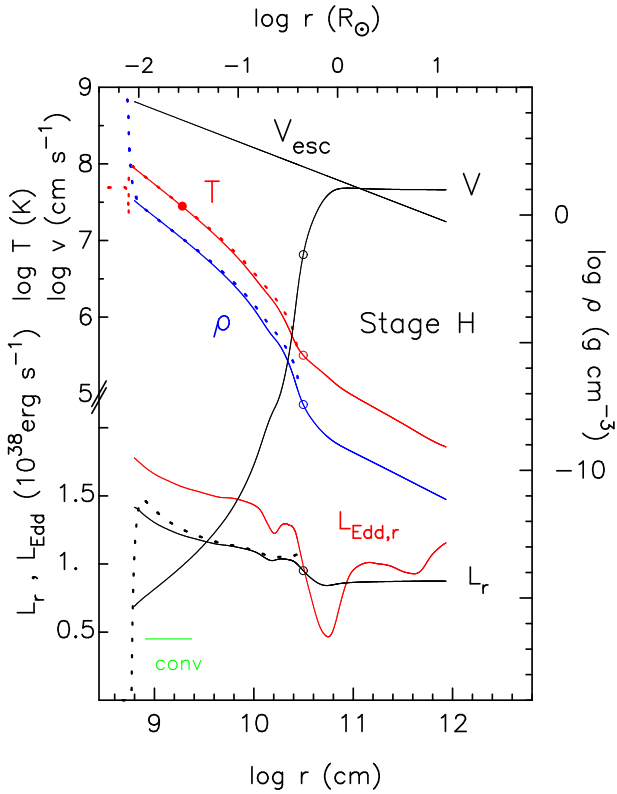}
  \end{center}
\caption{Same as in figure \ref{m10.struc.61131} but for stage H. 
}\label{m10.struc.64801}
\end{figure}

\begin{figure}
  \begin{center}
  \includegraphics[width=8.4cm]{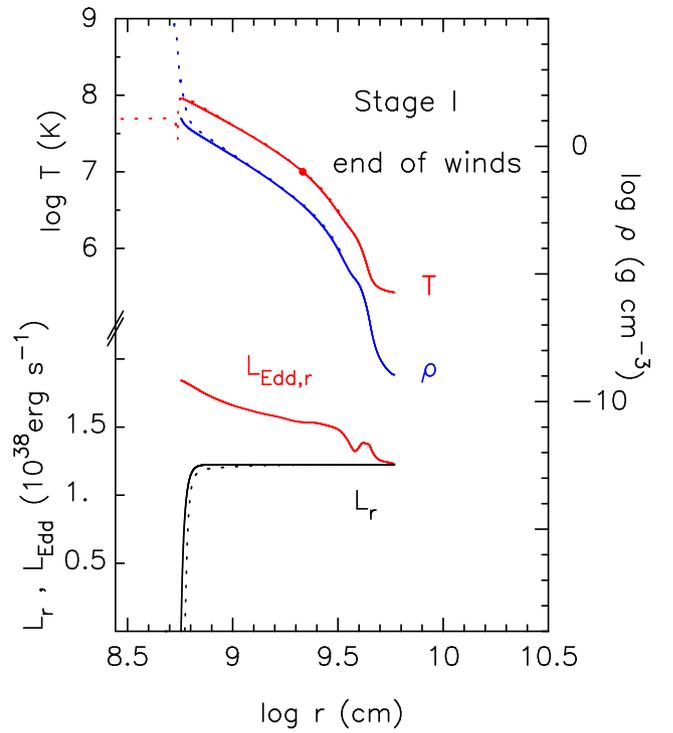}
  \end{center}
\caption{Same as in figure \ref{m10.struc.61131} but for stage I.
The optically thick winds have just stopped. 
}\label{m10.struc.MLlsmax2}
\end{figure}

After stage G the photospheric temperature (radius) rises (decreases)
with time. 
The wind mass loss rate turns to decrease (figure \ref{dmdt}). 
The mass of the hydrogen-rich envelope decreases 
owing mainly to wind mass loss and secondarily to nuclear burning. 
Figures \ref{m10.struc.64801} and \ref{m10.struc.MLlsmax2} 
show the internal structures at stage H and stage I (wind stopping), 
respectively.   
Both the figures show the inner Henyey code solution (dotted line parts) and 
the steady-state wind solution (solid line parts).  The fitting point
between the two solutions is indicated by a filled circle
on the temperature line.      
In the leftside to the fitting point we adopt the Henyey code solution,
while the steady state wind solution in the exterior
to the fitting point. 
In the both sides the dotted line is very close to the solid line because  
the envelope has settled to steady state. 
In stage I, throughout the envelope is in hydrostatic
because the wind has just stopped.  The structure is very close to
that of the hydrostatic (Henyey code) solution. 
In this way, the envelope evolves almost keeping steady state structure 
in the wind phase.

\begin{figure}
  \begin{center}
  \includegraphics[width=7cm]{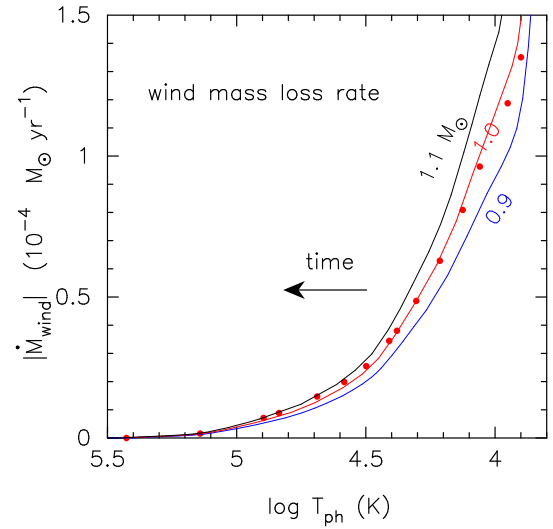}
  \end{center}
\caption{The wind mass loss rate vs photospheric temperature.
The dots indicate those of the present work (optically thick wind solutions).
The solid lines are the mass loss rates obtained by \citet{kat94h} 
for three WD masses of $1.1 ~M_\odot$ (black line), $1.0 ~M_\odot$ (red line), 
and $0.9~M_\odot$ (blue line) with solar composition. 
}\label{dmdtT}
\end{figure}

Figure \ref{dmdtT} shows the mass loss rates after the optical peak 
obtained in this work in the $T_{\rm ph}$ - $\dot M_{\rm wind}$ diagram. 
The right-top dot corresponds to stage G and the left-bottom dot 
to stage I ($\dot M_{\rm wind}=0.$). 
The mass loss rate decreases as $T_{\rm ph}$ increases with time 
(see figure \ref{dmdt} for the time dependence).  
The solid lines show the wind mass loss rates in the steady-state
sequence \citep{kat94h}. 
Our evolution model (dots) shows good agreement with the $1.0~M_\odot$ line. 
This will be discussed in Section \ref{sec_steadystate}.

\subsection{Envelope expansion and wind mass loss}
 
\begin{figure*}
  \begin{center}
 \includegraphics[width=12cm]{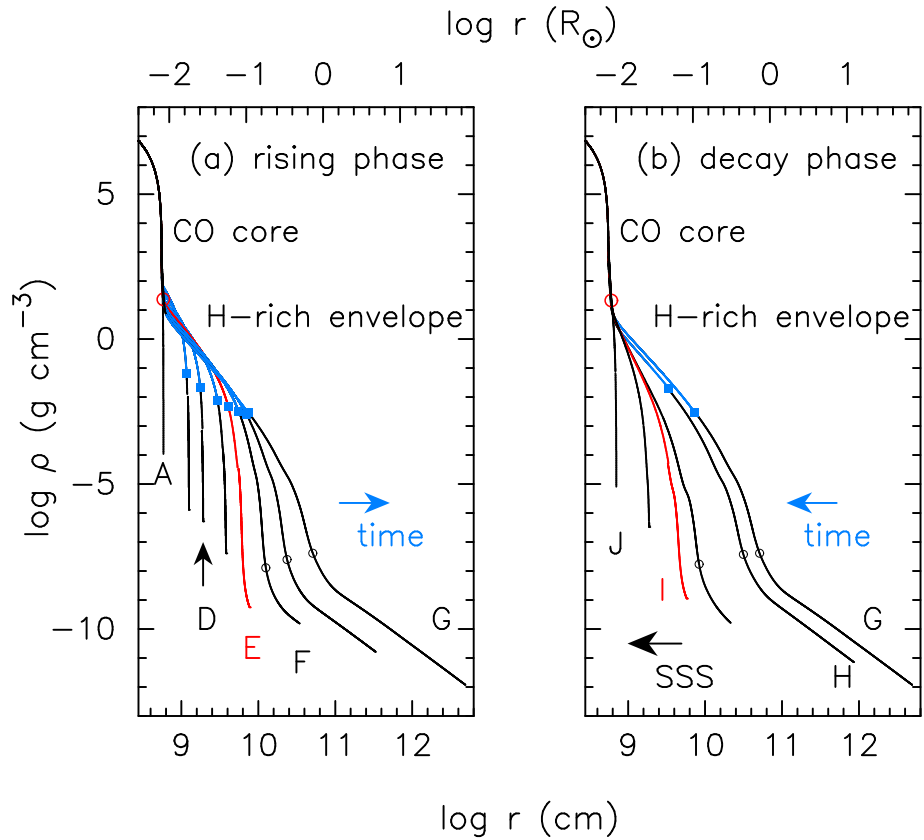}
  \end{center}
\caption{Temporal change in the density distribution:  
(a) in the rising phase from stage A to stage G,
(b) in the decay phase from stage G to stage J.  
The designated stages are indicated under the lines. 
The convective region of each stage is plotted in blue 
except stage E, and the filled blue squares indicate their 
outer edges, i.e., the border between the convective and radiative 
regions. The wind occurs at stage E and stops at stage I, both 
of which are indicated by the red lines. 
The critical points of the wind solutions are indicated by 
the small open black circles. The large open red circles show the places 
of maximum energy generation rate of nuclear burning 
$\epsilon_{\rm nuc}^{\rm max}$, for stages E and I, which is close to
but slightly above the bottom of the H-rich envelope. 
}\label{rho.2}
\end{figure*}

Figure \ref{rho.2} shows the temporal change of 
the density distribution. 
Before the outburst (stage A), the envelope is in plane parallel
structure where the density steeply decreases outward. 
After the shell flash starts the convection widely extends 
above the nuclear burning zone. 
In the early phase, such as stage D, the density slowly decreases 
outward in the convective region, keeping a steep gradient in the 
radiative region. 
This means that the envelope expansion in this early stage 
occurs in the convective region. 
The convection carries thermal energy upward.  As a result, 
the whole convective envelope is heated up and expands. 

After stage E the envelope expands faster in the outer region. 
The photospheric temperature decreases enough for 
the opacity to increase outward. 
Then the radiation pressure gradient increases.
If the gradient is large enough for the outer region of envelope
to be accelerated to supersonic and eventually to the escape 
velocity, such matter leaves as the optically thick winds. 
In the outer region above the critical point, the density decreases as 
$\rho \propto r^{-2}$ because of the steady state condition 
$4 \pi r^2 \rho v=$ constant with a constant velocity 
(see figures \ref{m10.struc.61161} to \ref{m10.struc.64801}). 
To summarize, the envelope starts its expansion first
in the convective region, followed by the entire envelope expansion,
but especially in the outer radiative region.  The transition from the 
subsonic expansion to radiative-driven supersonic wind is smooth. 

Panel (b) in Figure \ref{rho.2} shows the decay phase. 
Convection still exists at stages G and H, but disappears after stage H. 
The photospheric radius shrinks with time 
because the density in the outer envelope decreases.
After stage I, the wind has stopped.
The envelope further shrinks as the envelope mass decreases owing to
nuclear burning, and eventually becomes geometrically thin. 
At stage J the structure is in plane parallel.


\section{Energy budget} \label{sec_energy}

In this section we discuss how much amount of nuclear energy 
is consumed to drive the wind mass-loss. 

\subsection{Energy emitted from the photosphere}

First of all, we consider a simplest case in which 
no mass loss occurs during the outburst. 
Before the onset of thermonuclear runaway, the envelope is cold and 
geometrically thin. 
When a shell flash begins the envelope becomes hot and then expands. 
Thus, the generated nuclear energy is converted to (1) thermal energy 
to heat the envelope matter, (2) gravitational energy to expand the envelope 
against the gravity, and 
(3) radiative energy that is emitted from the photosphere, 
(4) kinetic energy of expanding matter.
Also some of the nuclear energy may flow inward to (5) heat the WD interior, 
but this inward flow is negligibly small 
because we adopt such a warm WD model that the 
WD is in heat balance with the long-term accretion.
Thus, the net flux into the WD interior is almost zero 
(see discussion in \citet{kat171500}).    

After the envelope expands to the maximum size, the envelope begins to 
shrink. The gravitational energy is converted to 
thermal energy, and eventually emitted from the photosphere. 
The envelope becomes cool and geometrically thin. 
In short, the produced nuclear energy are all 
emitted from the photosphere i.e., 
$\int L_{\rm nuc} dt= \int L_{\rm ph} dt$.

\subsection{Nuclear energy lost in the winds} 

\begin{figure}
  \includegraphics[width=8cm]{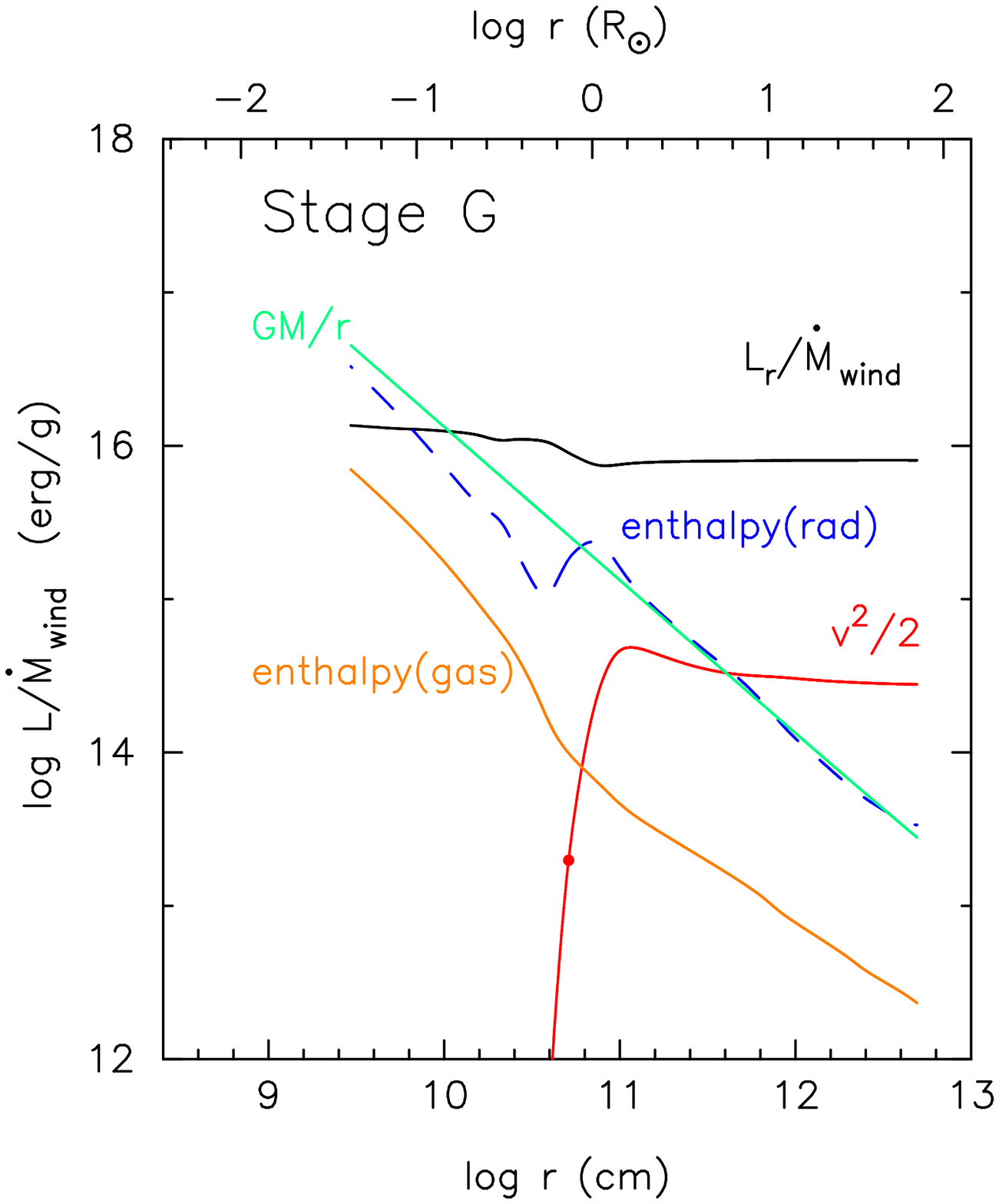}
\caption{The distribution of the diffusive radiative luminosity 
per unit mass, $L_r/\dot M_{\rm wind}$, and energy advected 
by winds at the maximum expansion of 
the photosphere (stage G): 
 gravitational energy $GM/r$, 
radiation enthalpy (rad), gas enthalpy (gas), 
and kinetic energy $v^2/2$. 
The filled circle on the kinetic energy denotes the critical point. 
The structure is plotted only for the region outside to the 
fitting point, i.e., the steady-state wind part of envelope. 
}\label{ent}
\end{figure}

When mass loss occurs, the ejected matter carries energy away from 
the WD. In the steady state wind the energy conversion is expressed by
an integrated form of \citep{kat94h}
\begin{equation}
L_r-\dot M_{\rm wind} ({v^2\over 2}+\omega_{\rm rad}+ 
\omega_{\rm gas}-{G M\over r}) = \Lambda,  
\label{equation_Lr}
\end{equation}
where $\Lambda$ is a constant throughout the envelope (strictly speaking,
outside the fitting point) at a given time $t$,
and  $L_r$ is the local diffusive luminosity.
The other terms are the energies carried with the moving matter, that is, 
the kinetic energy, enthalpy of radiation $\omega_{\rm rad}$ 
(photon energy trapped in the moving matter), 
enthalpy of gas $\omega_{\rm gas}$, and  
gravitational energy. 

We depict the distribution of each energy advected by winds 
and radiative luminosity 
at the stage of maximum expansion (stage G) in figure \ref{ent}. 
In the deep interior of envelope the advection of 
radiation energy (enthalpy of radiation) 
is dominant, which decreases outward because it is used to
lift up the envelope matter against the gravity.
Thus, the both energies (enthalpy of radiation and gravitational energy)
are compensated with each other. 
As a result, the diffusive luminosity $L_r/\dot M_{\rm wind}$ is dominant 
near the photosphere.  

\subsection{Energy budget throughout the outburst} 

\begin{longtable}{*{7}{c}}
 \caption{Energy budget}\label{table.wind}
  \hline              
    subject  &$E_{\rm nuc}$&$M_{\rm ej}$ & $E_{\rm rad}$&$E_{\rm G}$&$E_{\rm kin}$  \\ 
   unit  &(erg)&($M_\odot$) &(erg) &(erg) & (erg) \\
\endfirsthead
  \hline
  Name & Value1 & Value2 & Value3 \\
\endhead
  \hline
\endfoot
  \hline
\endlastfoot
  \hline
    &$3.2\times 10^{46}$ &$3.0\times 10^{-5}$ &$2.1\times 10^{46}$ & $1.1\times 10^{46}$  &$7.3 \times 10^{43}$ \\
\% &  100 &-- & 64 &35 &0.23 \\
\end{longtable}

The nuclear energy released during the outburst is 
calculated to be $E_{\rm nuc}=\int L_{\rm nuc} dt=3.2 \times 10^{46}$ erg 
from the beginning of the outburst until the epoch of accretion restart. 
The energy radiated from the photosphere in the same period is 
$E_{\rm ph}=\int L_{\rm ph} dt =2.06 \times 10^{46}$ erg.
This corresponds to 64 \% of the total nuclear energy generation. 
The rest of the energy is lost in the winds, mainly in the 
gravitational energy of ejecta
$E_{\rm G}=\int GM_{\rm WD}\dot M_{\rm wind}/R_0 dt
=1.1\times 10^{46}$ erg (35 \%), here $R_0=9.95\times 10^{-3}R_\odot$ 
is the radius originally located before the outburst. 
The kinetic energy of the ejecta is small 
$E_{\rm kin}=\int (1/2) \dot M_{\rm wind} v^2 dt =7.3 \times 10^{43}$ erg, 
corresponding only to 0.23 \% of the nuclear energy. These values are
summarized in Table \ref{table.wind}.

\section{UV Flash and X-ray flash}\label{sec_UV}

\begin{figure}
  \includegraphics[width=8.4cm]{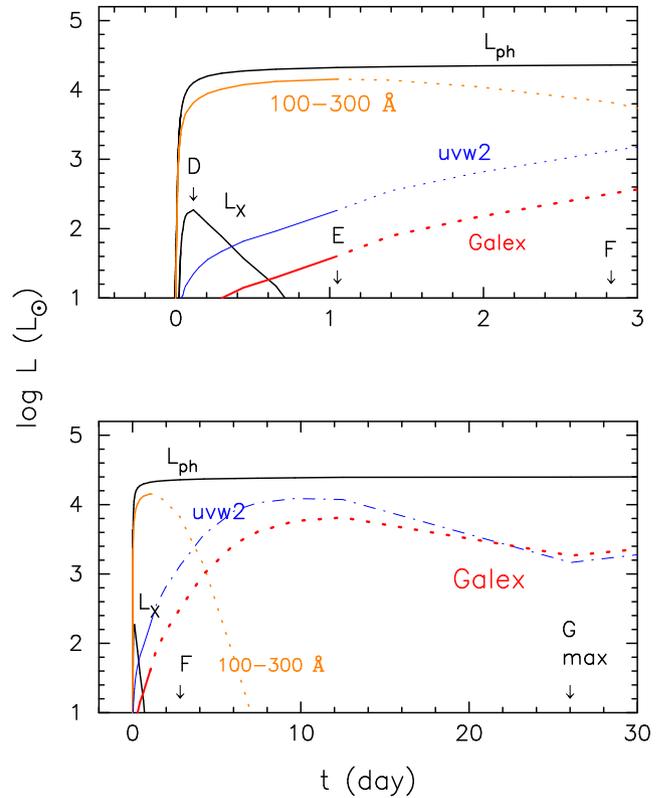}
\caption{The photospheric bolometric luminosity, $L_{\rm ph}$ 
(upper black line in each panel), luminosities of the supersoft X-ray 
$L_{\rm X}$ ($0.3 - 1.0$ keV: lower black line), 
$100-300$ \AA~band (orange line), uvw2 ($1120 - 2640$ \AA) band 
(blue line), and the Galex UV ($1786 - 2846$ \AA) band (red line). 
The upper panel shows the light curves for the 
first three days of the outburst, while lower shows the first 30 days. 
The stages D, E, F, and G are indicated. 
The supersoft X-ray luminosity reaches its maximum at stage D. 
The wind mass loss starts at stage E, 
so these high energy fluxes of the supersoft X-ray, $100-300$ \AA, 
uvw2, Galex UV bands could be partly self-absorbed by the wind itself. 
Stage G corresponds to the stage of the maximum 
wind mass-loss rate and possibly the optical peak.
}\label{xflash}
\end{figure}

After a thermonuclear runaway sets in, the WD immediately brightens 
up and its luminosity approaches the Eddington luminosity. 
The photospheric temperature is as high as to emit far UV flux. 
We plot the multiwavelength light curves for the first few days 
in figure \ref{xflash}, that is, supersoft X-ray 
($0.3 - 1.0$ keV: $12.4 - 42.3$ \AA), 
far UV ($100-300$ \AA), and UV ($1120 - 2640$ \AA~ for the Swift uvw2 band),
assuming black body emission. 
In the first day, the WD is very bright in the $100-300$ \AA~ band, 
but much fainter in the other bands.  
In the following days, these fluxes may be substantially reduced 
because the optically thick winds start at stage E and 
may absorb high energy photons. 
Thus, the nova is very bright in the $100-300$ \AA~ band 
only in the first day but may soon become fainter. 
This is the far-UV flash. In our model, it occurs 25 d before the optical peak.
Our calculation, however, does not include the effects of
pre/post-outburst mixing between the WD interior matter and
the accreted matter. If an effective mixing occurs, it increases 
the amount of heavy elements and the evolution timescale may be changed.  
This mixing effect is discussed in subsection \ref{sec_composition}.

The flux of each band strongly depends on the WD mass. 
In more massive WDs the temperature $T_{\rm ph}^{\rm max}$ 
is higher so that the X-ray flux is stronger and the $100-300$ \AA~
band flux is relatively weaker.  The X-ray flash light curve is presented 
by \citet{kat16xflash} for $M_{\rm WD} \ge 1.35~M_\odot$
and by \citet{kat17} for $M_{\rm WD}= 1.2~M_\odot$ and 1.38 $M_\odot$.  
For $M_{\rm WD}= 1.38~M_\odot$ the X-ray flash lasts about 1 day
and is as bright as $\log (L_{\rm X}/L_\odot)  \sim 4.4$
because the temperature $\log T_{\rm ph}^{\rm max}$ (K) $\sim 6.1$ 
is much higher than that of the 1.0 $M_\odot$ WD. 

\citet{mor16} and \citet{kat16xflash} attempted to detect 
such early X-ray flushes, but not detected.\footnote{ 
\citet{kon22wa} reported 
an X-ray flash in the classical nova YZ Ret 
on 2020 June 26 
with the eROSITA instrument on board Spectrum-Roentgen-Gamma (STG). 
This is the first detection of X-ray flash in novae.    }
\citet{cao12} presented multiwavelength light curves of 29 novae in M31.  
Their light curves include optical, IR, and GALEX UV 
(Galaxy Evolution Explorer satellite: effective wavelength 2316 \AA~with the
band width of 1060 \AA) bands.
They detected UV fluxes in few novae prior to the apparent optical maximum. 
For comparison, we plot the theoretical UV light curve in Figure \ref{xflash}, 
calculated from the blackbody luminosity 
between 740 \AA~ and 1703 \AA.

%


\section{Discussion}\label{sec_discussion}

\subsection{Comparison with steady state sequence model}
\label{sec_steadystate}

\begin{figure*}
  \begin{center}
  \includegraphics[width=12cm]{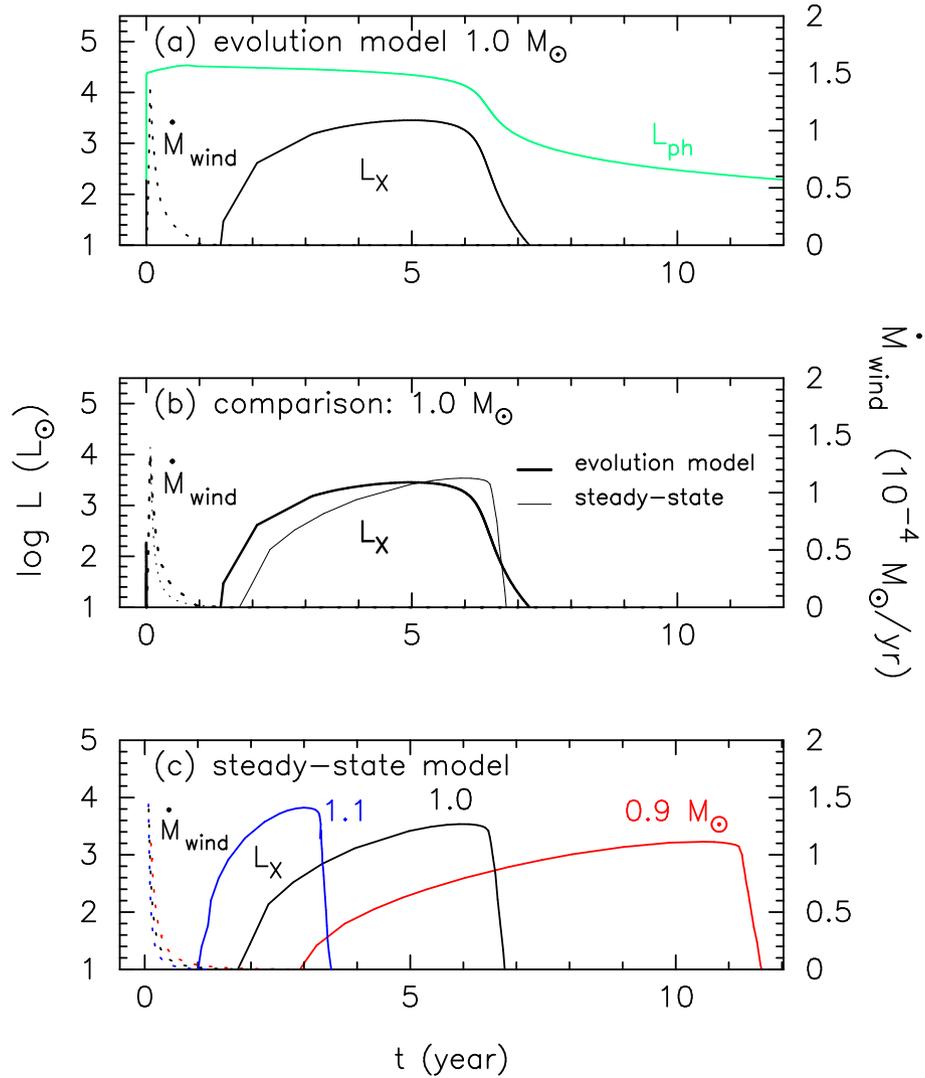}
  \end{center}
\caption{
Comparison of the X-ray light curves $L_{\rm X}$ ($0.3-1.0$ keV)
between our evolution model and steady state sequence.
(a) Evolution model of our 1.0 $M_\odot$ WD.
The green line shows the photospheric luminosity ($L_{\rm ph}$), 
whereas the solid black line is the X-ray light curve
($L_X$, scale on the left-side ordinate).
The X-ray flux is prominent in the X-ray flash ($t \sim 0$) and 
SSS phase (from $\sim 2$ yr to $\sim 6$ yr). 
The dotted black line indicates the wind mass loss rate with the scale 
on the right-side ordinate.  
(b) Comparison of our evolution model (in panel (a)) with 
the steady-state decay phase model of $1.0 ~M_\odot$ (thin lines). 
The origin of the steady-state model is set at 
the peak of the wind mass loss in the evolution model. 
(c) Three steady state sequence models of $0.9 ~M_\odot$ (red line),
$1.0 ~M_\odot$ (black), and $1.1 ~M_\odot$ (blue) WDs with the 
solar abundance \citep{kat94h}.
The $1.0 ~M_\odot$ WD model is the same as that in panel (b). 
}\label{light.theory2}
\end{figure*}

\citet{pri86} calculated one cycle of nova outburst
on a $1.25~M_\odot$ WD and showed that the envelope structure at $t=10^6$ sec 
is quite close to steady state over a wide region, 
from the inside to the critical point ($r_{\rm cr}=3.4~R_\odot$) to the 
outside of the photosphere ($r_{\rm ph}=36~R_\odot$). 
\citet{kat83} proposed a new way to follow the decay phase of nova outbursts 
with the sequence of steady-state solutions of decreasing envelope mass. 
This model is based on the assumption that the envelope has been 
settled down to steady state after the optical peak. 

In 1990's the opacity tables are revised. The new opacities 
(OPAL or OP: \citet{sea94}) has 
a prominent peak at $\log T$ (K) $ \sim 5.2$ owing to Fe ionization, 
even for the solar abundance. This large peak causes 
strong acceleration of winds and envelope structure during 
an extended phase has been drastically changed \citep{kat94h}. 
The evolution calculation becomes much difficult, and thus, 
each group has adopted each mass loss scheme to avoid 
the opacity peak and to accomplish their calculation until the end 
of a flash. 
In such treatments the photospheric temperature hardly 
decreases much below $\log T$ (K) $ \sim 5.2$, so 
it is difficult to calculate theoretical light curves 
to be compared with actual novae of which 
temperature typically decreases to $T_{\rm eff} < 10,000$ K 
(e.g., V339 Del: $T_{\rm ph} \ltsim 10,000$ and $R > 100~R_\odot$ \citet{sko14}). 

Now, the optically thick wind theory is only the way to properly calculate 
nova light curves, at least, in the wind phase.
Until now we have presented two different ways for 
calculating nova light curves based on the optically thick wind theory,
that is, 
(i) quasi-evolution sequence of steady state solutions, 
assuming the entire envelope is in steady state \citep{kat94h},  
and (ii) evolution calculation with Henyey code, in which 
steady state solutions are incorporated in the wind phase (this work). 

The first method (i) has been studied and applied to a number of novae. 
Kato and Hachisu have systematically calculated steady state sequences 
with various parameters for classical novae 
and have fitted observational properties of various types of novae 
\citep{hac06kb, hac07k, hac08kc,hac10,hac14ka,hac15k,
hac16a,hac16b,kat09b,kat21,kat20sh}.

The second way (ii) is the one we took the present work. 
This is the first accurate time-dependent calculation for classical novae, 
so that we have a chance to compare these two ((i) and (ii)) evolution sequences.
The treatment (i) is less accurate because it lacks some 
time-dependent terms, but the calculation is much easier than that of (ii).
We already confirmed in section \ref{sec_structure} 
that the internal structure of the envelope 
is close to that of the steady state solution in the decay phase. 
In what follows we compare the evolution timescale of these two nova
models.

Figure \ref{light.theory2}(a) shows our evolution model (i). The solid 
lines denote the bolometric luminosity $L_{\rm ph}$ and the supersoft 
X-ray flux $L_{\rm X}$. The X-ray flash is shown at $t \sim 0$ yr 
and the SSS (supersoft X-ray source) phase emerges from $t\sim 2$ yr to 
$\sim 6$ yr.  We plot the wind mass loss rate (dotted line)
instead of the optical magnitude because the optical flux in most novae 
are dominated by free-free emission which is closely related to the 
wind mass loss rate. 

Figure \ref{light.theory2}(b) compares our evolution model 
(thick line) with a steady-state sequence model (thin line) 
for a $1.0~M_\odot$ WD with the solar composition 
($X=0.7$, $Y=0.28$, and $Z=0.02$)\citep{kat94h}. 
The steady state sequence model begins at $t=26$ day =0.07 yr, 
to fit stage G in the evolution model. 

Despite the difference between the two methods ((i) and (ii)),
the two X-ray light curves are in good agreement, in 
the main characteristic properties, i.e., the decay of wind 
mass loss rate, X-ray turn-on time, peak X-ray flux, and 
X-ray turn-off time. 
The main difference between the two models is 
the gravitational energy release $L_{\rm G}$. 
In the steady state model (i), $L_{\rm G}$ is partly included 
in the wind phase, but not included in the SSS phase. 
The resemblance of two different light curves 
is mainly due to small contribution of $L_{\rm G}$ in the evolution model (ii), 
which is a twentieth of $L_{\rm ph}$ in the SSS phase 
as shown in figure \ref{evol}.

The X-ray turn-off time and SSS duration 
are good indicators of the WD mass 
and have been used to estimate the WD mass of individual novae.
To see the dependence of a SSS phase on the WD mass, 
we plot panel (c) in Figure \ref{light.theory2}
for three steady state sequence models of different WD masses, i.e.,
$0.9 ~M_\odot$, $1.0 ~M_\odot$, and $1.1 ~M_\odot$ \citep{kat94h}. 
The light curves of the 0.9 and 1.1 $M_\odot$ show much different 
timescales and peak fluxes.  We may safely use the results of 
steady state sequence models (i) to estimate the WD mass. 

Whereas the Henyey code model (ii) needs a complicated process and many manual
iteration to calculate one cycle of a nova outburst,
calculation with steady state (wind/no wind) sequence (i) is 
relatively easier.  In this sense the steady state sequence (i)
is a good way to calculate the decay phase of a nova outburst.

\subsection{Effects of CO enrichment}\label{sec_composition}

During a nova outburst nuclear burning converts hydrogen to helium
but the total amount of heavy elements is unchanged.
In our 1.0 $M_\odot$ model the chemical composition of ejecta
changes from $X=0.7$, $Y=0.28$, and $Z=0.02$ (accreted matter)
to $X=0.65$, $Y=0.33$, and $Z=0.02$ (ejecta in the decay phase).
On the contrary, classical nova spectra show enhancements of 
heavy elements such as C, N, O, Ne, or Mg 
(see, e.g., \citet{hac10} for a summary),
which are considered to originate from WD interiors.

In spherically symmetric (1D) calculations such as ours, the convection
does not reach down to the CO core below the interface of the
accreted hydrogen-rich matter,
which prevents CO-rich matter from entering into the 
hydrogen-rich envelope. 
We did not include particular mixing mechanisms
of dredging up WD core material into envelope
to avoid free parameters such as an uncertain mixing degree.
Also, we did not include diffusion processes of nuclei in the
quiescent phase. The timescale of diffusion is estimated
to be $3.6\times 10^4$ yr, using Equation (A3) in \citet{she09}, 
for hydrogen atoms to diffuse from the bottom
of hydrogen-rich envelope ($X=0.1$) down to the top of He layer
($X=0.0003$).
This timescale is much longer than the accreting phase of 5358 yr,
so the diffusion process is not efficient.

Effects of CO enhancement on nova outbursts 
has been studied by \citet{sal05,den13,che19,sta20}.  
\citet{den13} and \citet{che19} reported that 
nova outbursts with CO enhancements in the accreted matter  
shows a smaller ignition mass, lower maximum temperature 
$T^{\rm max}$ at the shell flash, and less extent of excursion toward 
the lower temperature in the H-R diagram. 
From these calculations we see two opposite effects of CO enrichment. 
A larger CO content enhances the nuclear burning, which would increase 
the expansion rate of the burning shell and shorten the timescale of a
flash, that is, the X-ray flash phase would become shorter and the wind 
mass loss would start earlier, if the other parameters are the same.  
On the other hand, the flash will be weaker for a smaller ignition mass, 
which would increase the timescale of a flash. 
These two effects work in the opposite direction, and it is not clear, 
at present, which effect is more important in the early phase of a 
nova outburst. 

For the decay phase, on the other hand,  
the effects of heavy element content have been systematically 
studied with the steady state sequence models
\citep{kat97,hac06kb,hac10,hac15k,hac16a,kat21}.  
For a larger $Z$ both the wind mass loss rate and velocity increase. 
The light curves decay faster after the optical maximum, 
and the following supersoft X-ray source phase becomes shorter.
\citet{hac06kb} obtained the wind durations $t_{\rm wind}= 218$ days
for CO nova 2 (CO2) chemical composition 
($X=0.35$, $Y=0.33$, $X_{\rm CNO}=0.30$, and $Z=0.02$)
and  $t_{\rm wind}= 712$ days for solar ($X=0.70$, $Y=0.28$, and $Z=0.02$),
where the WD mass is $1.0~M_\odot$.  The timescale of CO2 composition
is about 3 times shorter than that of solar. 

In our present 1.0 $M_\odot$ model the time from the
start of ignition to the maximum expansion of the photosphere is about
26 days (see Table \ref{table.time}).  On the other hand, a typical
timescale from the ignition to the maximum brightness is estimated 
as from a few to several days
(e.g., 3 days in V339~Del, see \citet{schaefer14bg} and \citet{sko14}).
The difference may come from the difference of physical parameters such as 
abundance and mass accretion rate.

{\subsection{Comparison with Hillman et al. (2014)}
\label{sec_discussion.hillman}

Theoretical light curves of classical nova outbursts have rarely 
been  published because of the difficulties in calculating
a full cycle of a classical nova evolution.
\citet{hil14} presented photospheric 
temperature variations and  multiwavelength light curves 
based on the hydrogen shell flash models 
obtained by \citet{pri95} and \citet{yar05}. 
In this subsection we compare our 1.0 $M_\odot$ 
classical nova model with Hillman et al.'s 100.10.8 model
($M_{\rm WD}=1M_\odot$, $T_{\rm c}=10^7$K, 
$\dot{M}_{\rm acc}=10^{-8}\,M_\odot$yr$^{-1}$).

\begin{figure*}
  \begin{center}
  \includegraphics[width=10cm]{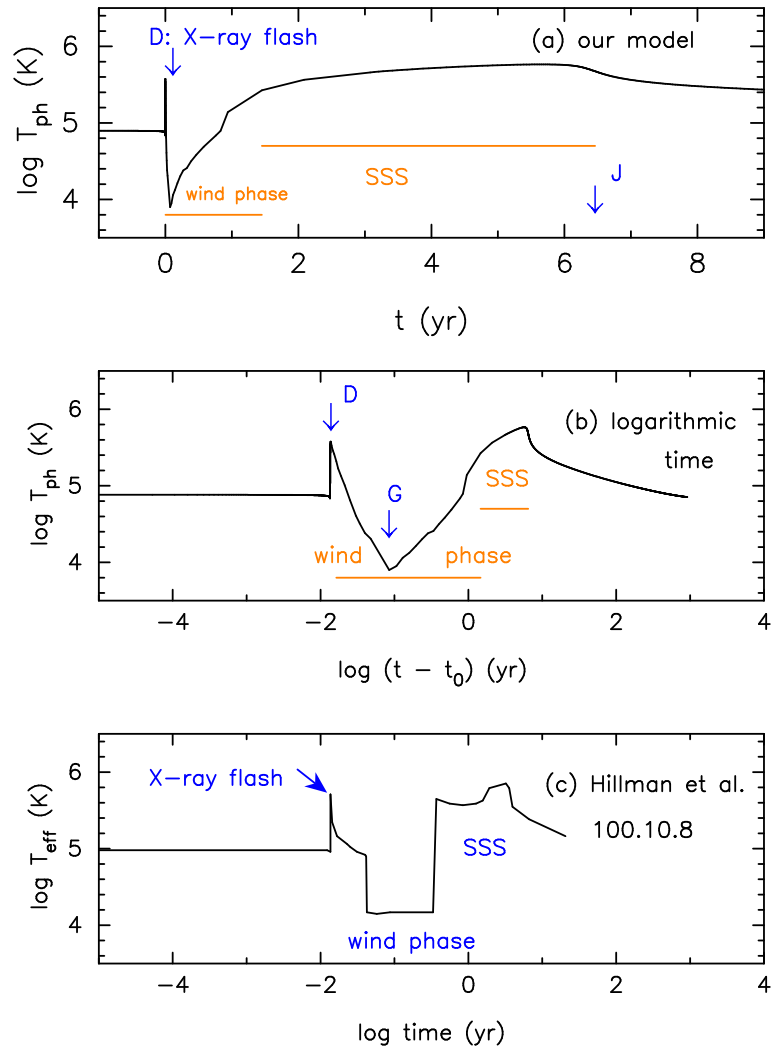}
  \end{center}
\caption{Comparison of the temperature evolution of 
our model with \citet{hil14}'s model. 
(a) Our photospheric temperature is plotted against time, where  
$t=0$ is the time at $L_{\rm nuc}^{\rm max}$.  
(b) The same quantity as in panel (a) is plotted as a function of a 
logarithmic time, in which the origin of time is shifted by   
4.93 days before stage B to make comparison with panel (c) easier.  
(c) The temporal change of $T_{\rm eff}$ in model 100.10.8 
($1~M_\odot, T_{\rm c}=10^7$ K, $\dot M_{\rm acc}=10^{-8}~M_\odot$yr$^{-1}$)
of \citet{hil14}. 
}\label{T}
\end{figure*}

\subsubsection{Photospheric temperature}

Figure \ref{T}(a) shows the 
photospheric temperature $T_{\rm ph}$ variation 
of our model as a function of time. 
The wind phase is the period between stage E and stage I, and 
the SSS phase is between stage I and stage J. 
To compare Hillman et al.'s temperature profile, 
we adopt a logarithmic time for the abscissa of panel 
(b), in which the origin of time is shifted by 4.93~days.

After the ignition of a thermonuclear runaway, 
$T_{\rm ph}$ quickly reaches maximum (stage D).
Then $T_{\rm ph}$ decreases as the envelope expands, and   
takes a minimum value in a few days.  After the maximum expansion
of the photosphere ($T_{\rm ph}$ minimum), $R_{\rm ph}$ begins to contract
with the photospheric temperature being increased as the mass of hydrogen-rich
envelope decreases mainly by wind mass loss.
When $T_{\rm ph}$ increases to $\sim 3\times10^5$~K, 
the nova enters a supersoft X-ray source (SSS) phase. 

Figure \ref{T}(b) and (c) show photospheric/effective temperature 
variations in a logarithmic time of ours and Hillman et al.'s, respectively.   
Although the amplitude of photospheric temperature 
variation of our model is comparable 
with that of Hillman et al., 
panel (b) shows a smooth decrease/increase 
in the cool (expanded) phase  
while panel (c) displays the sudden changes. 
The variation of photospheric temperature
in the expanded phase is correlated 
with wind mass-loss rates (see Figure \ref{dmdtT}). 
Such a sudden decrease/increase of $T_{\rm eff}$ should 
accompany some drastic change in the envelope,  
but no description is given in \citet{pri95}, \citet{hil14},
and \citet{yar05}. 

Observationally 
gradual temperature increases are reported after the optical maximum  
(\citet{cas02,cas04,kat09b}). These smooth temperature increases are 
consistent with our model.



\subsubsection{Multiwavelength light curves}

\begin{figure*}
  \begin{center}
  \includegraphics[width=10cm]{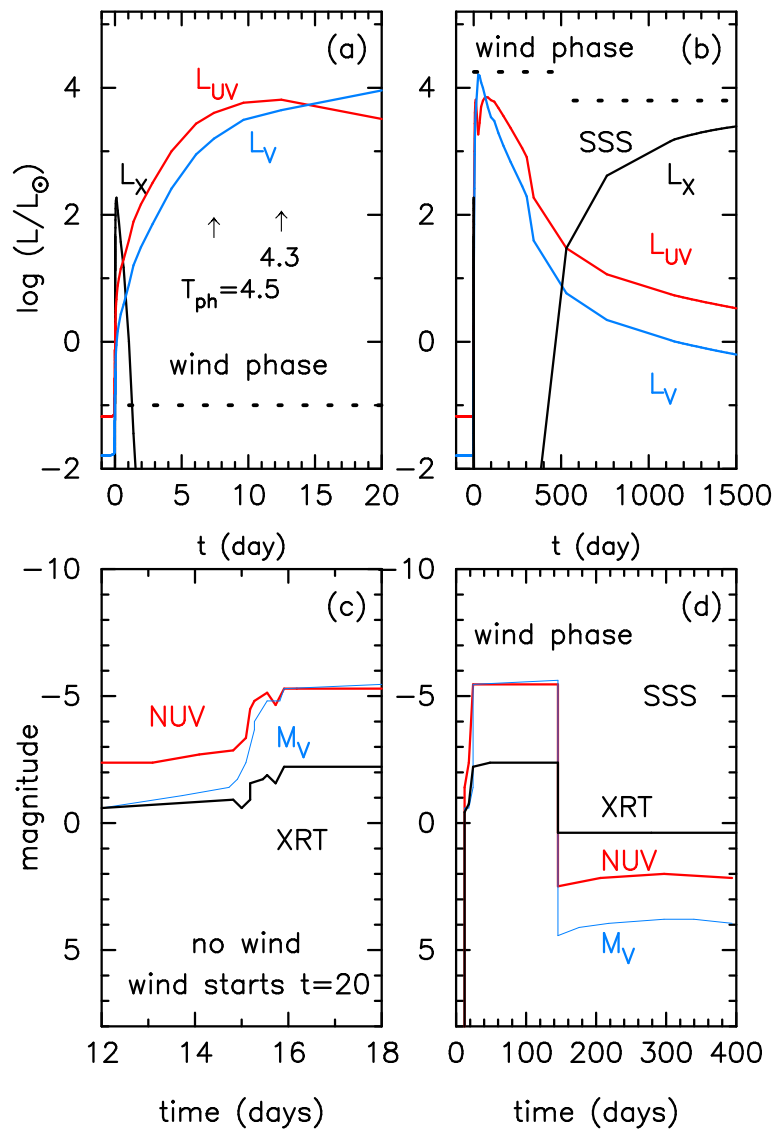}
  \end{center}
\caption{Comparison of light curves between our model and 
Hillman et al.'s model. 
The upper two panels show three wavelength bands in $V$, UV and 
X-rays in 20 days (a) and 1500 days (b). The wind phase 
and SSS phase are indicated by the dotted horizontal lines. 
In panel (b) the X-ray flash is overlapped to the other two lines 
at $t$ (day) $\sim 0$. 
The lower two panels show the model 100.10.8 
($1M_\odot$, $T_{\rm c}=10^7$K,~$\dot{M}_{\rm acc}=10^{-8}$,
$M_\odot$ yr$^{-1}$) of \citet{hil14}. 
}\label{hillman}
\end{figure*}

Figure \ref{hillman}(a) and (b) show our light curves of three 
wavelength bands, i.e, supersoft X-ray, Galex UV, and $V$
calculated from $T_{\rm ph}$ and $L_{\rm ph}$ assuming blackbody. 
Panel (a) shows the first 20 days, whereas panel (b) covers 1500 days. 
Note that the blackbody $V$ light curve $L_{\rm V}$ does not accurately
correspond to the observed ones because the emission in classical novae
is dominated by free-free emission not by blackbody emission
(see, e.g., Figure 4 in \citet{kat17palermo} for the comparison of 
the light curves of free-free emission and blackbody). 
In the early phase (in panel (a)) the brightest emission band shifts from
X-ray to UV toward the optical peak ($t=26$ day).  In the decay phase  
(panel (b)), it shifts from optical, UV to X-ray.  

The lower two panels show light curves in magnitudes, obtained by 
\citet{hil14} for visual ($M_V$), GALEX NUV, and soft X-ray (Swift XRT),
which are reproduced from their figure 4 (panel 2).
We did not find any definitions on their origin of time and
X-ray magnitude. 

Hillman et al.'s light curves show sudden changes,
which are related with the nearly discontinuous temperature decrease to 
and rise from the minimum $\log T_{\rm eff}$(K)$\approx 4.2$ 
as seen in Figure\,\ref{T}(c). 
These variations are stark contrast to the smooth 
changes of our luminosity and temperature curves in panels (a) and (b).
Our supersoft X-ray light curve shows a precursor flash 26 days 
before the $V$ maximum while a much stronger precursor flash
occurs in a far UV range of $100-300$~\AA~  (figure \ref{xflash}).  

\citet{hil14} claimed that they theoretically predicted  
the ``UV precursor flashes and pre-maximum halts.'' 
These features are different from our smooth light curves. 
No X-ray flash exists in panel (d). 
We strongly encourage the authors to publish internal structures 
of the envelope and mass loss rates 
(see discussion and Table 1 in \citet{kat17palermo}).

\section{Conclusions}\label{section_conclusion}
 
Our main results are summarized as follows.

\noindent
1. We calculated self-consistent models for one cycle of 
a classical nova outburst, 
in which winds are driven by the radiation pressure gradients. 
The model is a $1.0 ~M_\odot$ WD
with a mass accretion rate of 
$\dot M_{\rm acc}=5\times 10^{-9}~M_\odot$~yr$^{-1}$
(the recurrence period is 5370 years). 
The wind mass loss rate reaches a peak $1.4 \times 10^{-4}~M_\odot$~yr$^{-1}$
at the epoch of maximum expansion of the photosphere when the photospheric
temperature decreases to $\log T_{\rm ph}$ (K)=3.90.

\noindent
2.  The nuclear energy generated during the flash is emitted 
in a form of radiation (64 \%), or is converted to the 
gravitational energy of the ejecta by lifting them up against
the gravity (35 \%) and to the kinetic energy of the ejecta (0.23 \%). 

\noindent
3. The nuclear energy at thermonuclear runaway is once stored as 
heat energy ($L_{\rm G}$) and emitted gradually throughout
the entire period of the flash. 
Thus, no internal shocks or blast waves appear in the envelope. 
Also, the photospheric luminosity does not exceed
the Eddington luminosity at the photosphere. 

\noindent
4. In a very early phase of the outburst the 1.0 $M_\odot$ WD is very 
bright in a far UV band (100 - 300 \AA),
but not so bright in the supersoft X-ray band ($0.3-1.0$~keV).
This UV flash occurs 25 days before the optical peak.  It should be noted
that these individual characteristic properties depend very much on the
WD mass, mass accretion rate, and chemical composition of the envelope.

\noindent
5. The envelope structure approaches steady state 
after the maximum expansion of photosphere,   
mainly because the gravitational energy release $L_{\rm G}$ 
is negligibly small compared with diffusive luminosity $L_r$.
In such a circumstance (very small $L_{\rm G}$),
the optically thick wind theory (a sequence of steady state
wind solutions) is a good approximation to the decay phase of a nova outburst.

6. Hillman et al. (2014)'s three band light curves show a rectangle 
shape corresponding to the quick change of effective temperature. 
Our evolution model shows, on the contrarily, gradual change in the 
multiwavelength light curve, corresponding to the gradual change 
of structure, temperature, and wind mass loss rate.

\begin{ack}
We are grateful to the anonymous referee for useful comments that
improved the manuscript.
\end{ack}




%

\end{document}